\documentclass{Interspeech2024}
\usepackage{booktabs}
\usepackage{caption}
\usepackage{amsmath}
\usepackage{cite}
\usepackage{authblk}

\usepackage{hyperref}
\hypersetup{hidelinks,
	colorlinks=true,
	allcolors=black,
	pdfstartview=Fit,
	breaklinks=false}
\usepackage{times}
\usepackage{soul}
\usepackage[utf8]{inputenc}
\usepackage{caption}

\interspeechcameraready 

\title{SPA-SVC: Self-supervised Pitch Augmentation for Singing Voice Conversion}

\name{Bingsong Bai, Fengping Wang, Yingming Gao, Ya Li$^{*}$}

\address{
  School of Artificial Intelligence, Beijing University of Posts and Telecommunications, China
  }
\email{\{bingsongbai, wfp, yingming.gao, yli01\}@bupt.edu.cn}


\begin{document}
\maketitle
\begin{abstract}
 Diffusion-based singing voice conversion (SVC) models have shown better synthesis quality compared to traditional methods. However, in cross-domain SVC scenarios, where there is a significant disparity in pitch between the source and target voice domains, the models tend to generate audios with hoarseness, posing challenges in achieving high-quality vocal outputs. Therefore, in this paper, we propose a \textbf{S}elf-supervised \textbf{P}itch \textbf{A}ugmentation method for \textbf{S}inging \textbf{V}oice \textbf{C}onversion (SPA-SVC), which can enhance the voice quality in SVC tasks without requiring additional data or increasing model parameters. We innovatively introduce a cycle pitch shifting training strategy and Structural Similarity Index (SSIM) loss into our SVC model, effectively enhancing its performance. Experimental results on the public singing datasets
M4Singer indicate that our proposed method significantly improves model performance in both general SVC scenarios and particularly in cross-domain SVC scenarios. \footnote{Audio samples of our work can be found on the homepage below: \href{https://shawnpi233.github.io/publication/paper/spasvc}{https://shawnpi233.github.io/publication/paper/spasvc}}
\end{abstract}

\renewcommand{\thefootnote}{\fnsymbol{footnote}}

\footnotetext{Corresponding author.}
\renewcommand{\thefootnote}{\arabic{footnote}}
\section{Introduction}
Singing Voice Conversion (SVC) is a process where a piece of singing voice is transformed from one person's voice to another, while preserving the content. The essence of SVC lies in disentangling the timbre and content features within the singing voice and reconstructing the voice based on these disentangled features. In recent years, there has been rapid progress in self-supervised learning methods for speech encoding representations. Notable advancements include models like wav2vec 2.0 \cite{baevski2020wav2vec2}, HuBERT \cite{hsu2021hubert}, Contentvec \cite{qian2022contentvec}, and WavLM \cite{chen2022wavlm}. These efficient self-supervised speech encoding models provide robust support for disentangling timbre and content features within SVC models. Successful methods in this domain include state-of-the-art (SOTA) models like DDSP-SVC-Diff \cite{ddsp-svc2023}, So-VITS-SVC \cite{so-vits-svc2023}, DiffSVC \cite{liu2021diffsvc} and CoMoSVC \cite{lu2024comosvc}, among others. These models utilize the robust performance of self-supervised speech encoding models and the superior reconstruction capabilities of diffusion \cite{ho2020ddpm} and VITS \cite{kim2021conditional_vits} models, resulting in commendable performance in general SVC scenarios.

However, challenges persist in cross-domain SVC scenarios, where the pitch of the song melody exceeds the vocal range of the target domain singer. This often leads to voice distortions or complete loss of vocal quality, especially in cases where the pitch span is large. These limitations hinder the applicability of SVC techniques in scenarios requiring transformations across extensive vocal ranges. To the best of our knowledge, existing voice enhancement methods primarily focus on tasks such as accompaniment noise removal \cite{xu2023mbtfnet}, recording noise reduction \cite{liu2022voicefixer} and general voice enhancement within the typical vocal range \cite{guo2022singaug}. In current cross-domain SVC studies, FastSVC \cite{liu2021fastsvc_cross} aims for superior conversion performance and faster inference compared to standard systems, while HiFi-SVC \cite{zhou2022hifisvc_cross} prioritizes generating 22.05 kHz voices using advanced neural vocoders and convolutional modules for F0 modeling. Nonetheless, these models are limited to generating voices at 16 kHz or 22.05 kHz, potentially compromising audio quality in general SVC scenarios compared to higher sample rate (44.1 kHz) voices produced by the SVC models based on diffusion or VITS. Therefore, this paper focuses on exploring how to improve the model performance in both general SVC scenarios and particularly in cross-domain SVC scenarios.

Although existing SVC models can introduce small-scale pitch random perturbations during training, they often generate hoarse artifacts in cross-domain SVC scenarios, particularly as the vocal range span increases. Furthermore, in further experiments, we found that using SVC models to only increase the pitch of the source vocals without changing the timbre leads to hoarse audio, with hoarse portions persisting even when the pitch is lowered back to the original. Additionally, as the magnitude of pitch shifting increases, the range of hoarse artifacts expands, sometimes rendering the entire audio segment almost indistinguishable. When returning the pitch shifted hoarse voice to original pitch, hoarseness still remains.

\begin{figure}[b]
    \centering
        \includegraphics[width=\linewidth]{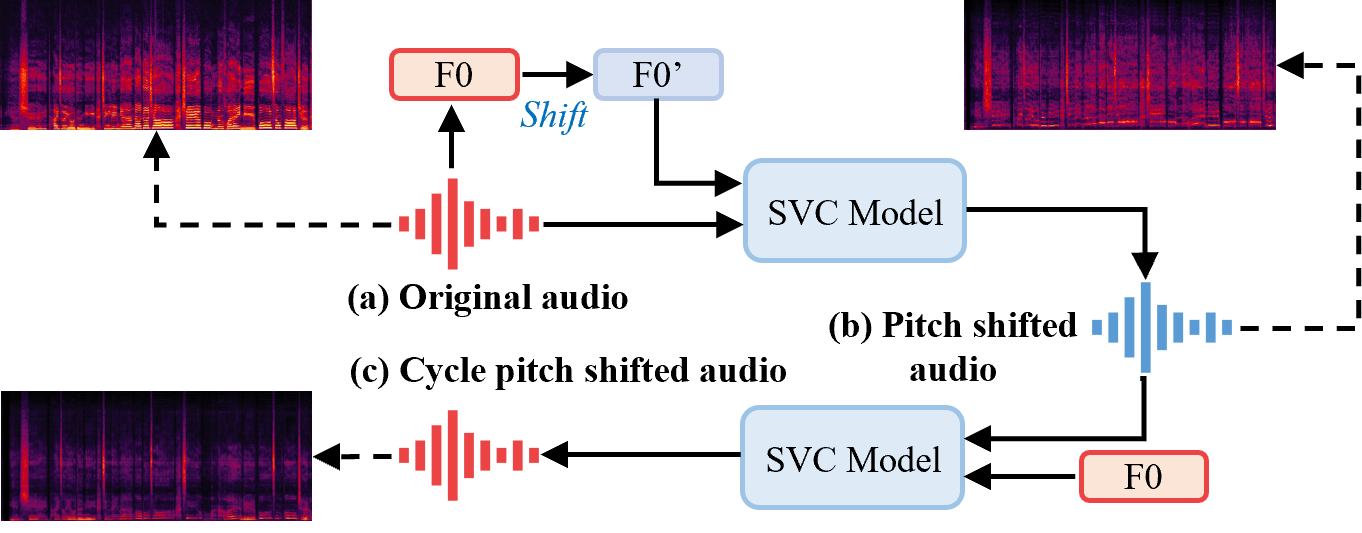}
        \caption{The pipeline of cycle pitch shifting.}
        \label{fig:audios}
\end{figure}

Therefore, we explored the possibility of cycle pitch shifting during model training. This involved introducing random perturbations within a certain range and then returning back to the original pitch, which could potentially produce partly hoarse audio alongside the original audio. This process is illustrated in Figure \ref{fig:audios}. By calculating the loss function of features between this hoarse audio and the original audio, we can optimize the model to reduce the occurrence of hoarse artifacts in SVC model generations. In addition, we found that the Structural Similarity Index (SSIM) \cite{wang2004ssim}, commonly used in image processing, can enhance the performance of SVC models when employed to compute the loss of mel-spectrograms. Due to its computational characteristics, SSIM enables the model to focus more on the overall structure and details of the data, thereby improving the overall performance of the SVC model.

Therefore, the main contributions of this work can be summarized as follows:
\begin{itemize}
\item Propose a \textbf{S}elf-supervised \textbf{P}itch \textbf{A}ugmentation method for \textbf{S}inging \textbf{V}oice \textbf{C}onversion called SPA-SVC. By integrating a cycle process of randomly raising and lowering pitch during model training, we generate hoarse voice aligned with the original voice. This can alleviate the problem of voiceless regions caused by large pitch spans.

\item Introduce SSIM loss into SVC and experimentally demonstrated its significant contribution to improving the quality of generated singing voices. 
\end{itemize}

The structure of this paper will be arranged as follows: Section \ref{sec:method} will present the methodology and Section \ref{sec:experiments} will describe the experimental setups and showcase the experimental results. And Section \ref{sec:conclusions} will conclude the paper.
\section{Method}
\label{sec:method}
This section will provide a detailed explanation of our training and inference architecture in SPA-SVC, as well as the modules utilized within the model. As illustrated in Figure \ref{fig:model}(a), we depict the architecture diagram for model training, while Figure \ref{fig:model}(b) showcases the architecture diagram for model inference.
\vspace{-0.2cm}
\subsection{SPA-SVC Architecture}
In line with SVC models based on self-supervised diffusion \cite{liu2021diffsvc, lu2024comosvc} or VITS \cite{zhou2023vitssvcc, peng2023singing_vits_opera}, the training process of SPA-SVC begins with the extraction of self-supervised learning features (SSL), volume features (Vol), fundamental frequency (F0), and, to distinguish between different speakers, a speaker ID determined based on the current speaker. It is noteworthy that, following the extraction of fundamental frequency, we apply a pitch shifting transformation (for clarity, we denote features from the original audio as ``o'' and highlight them in red, while features from pitch-shifted audio are denoted as ``s'' and highlighted in blue). Differing from minor-range pitch enhancement in other models \cite{guo2022singaug}, we set the number of the semitones in pitch shifting to a random integer within the range of 6 to 18, aiming to obtain a pitch-shifted singing voice with added hoarse voice during training. The formula for pitch augmentation is as follows:
\begin{figure}[t]
  \centering
  \includegraphics[width=\linewidth]{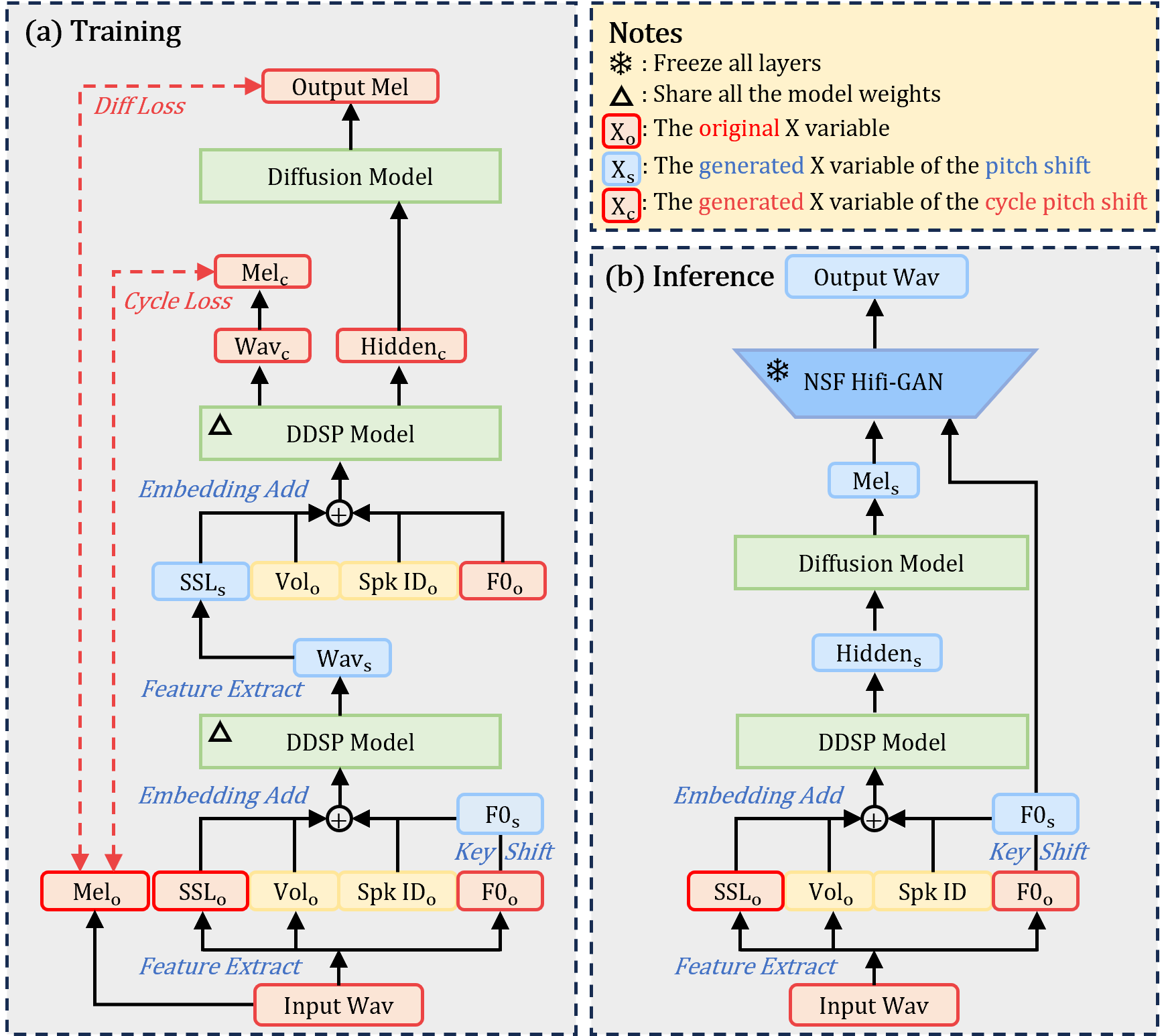}
  \caption{The training and inference architecture of SPA-SVC.}
  \label{fig:model}
\end{figure}
\vspace{-0.1cm}
\begin{equation}
F0_{s}= F0_{o} \times 2^{\left(\frac{key}{12}\right)},
\label{eq:1}
\end{equation}
where, $F0_s$ denotes shifted pitch, $F0_o$ denotes orginal pitch and $key$ denotes the number of semitones in pitch shifting. Afterward, $F0_s$ is combined with other original audio features and input into the DDSP model through an embedding layer. This process results in the first reconstruction, yielding a pitch-shifted singing voice $wav_s$ with added breathiness. Subsequently, the self-supervised learning features $SSL_s$ are extracted from $wav_s$, which, along with other original audio features, are jointly input into the same DDSP model for a second time. This cycle reconstruction produces a singing voice $wav_c$ with the original pitch restored. Consequently, by computing the loss between the mel-spectrogram of the cycle reconstructed audio and that of the original audio, we optimize the DDSP model to generate high-pitched singing voices less likely to contain hoarseness.

To investigate methods for enhancing the DDSP model's focus on finer details of singing voice and reducing the occurrence of hoarse sound, we introduce SSIM loss as cycle-consistency loss $\mathcal{L}_{cyc}$. The formula for SSIM loss is provided below:
\begin{equation}
\mathcal{L}_{cyc} = 1- \text{SSIM}(x, \hat{x}),\label{eq:2}
\end{equation}
\begin{equation}
\text{SSIM}(x, \hat{x}) = \frac{({2\mu_{x} \mu_{\hat{x}} + c_1})({2\sigma_{x\hat{x}} + c_2})}{({\mu_{x}^2 + \mu_{\hat{x}}^2 + c_1})({\sigma_{x}\sigma_{\hat{x}} + c_2})},\label{eq:3}
\end{equation}
where $x$ denotes mel-spectrogram of the original audio and $\hat{x}$ denotes mel-spectrogram of the cycle reconstructed audio from DDSP model. $\mu$ denotes mean value, $\sigma_{x\hat{x}}$ denotes covariance of $x$ and $\hat{x}$, $\sigma_{x}^2$ and $\sigma_{\hat{x}}^2$ denote the variance of $x$ and $\hat{x}$ respectively and $c_1$ and $c_2$ are two variables used to stabilize the division with a weak denominator.

Then, we utilize the output of the cycle reconstructed hidden layer features $hidden_c$ from DDSP as the input for the diffusion model. Subsequently, we obtain the final reconstruction of the original audio's mel-spectrogram $mel_c$, and compute the diffusion loss to train the diffusion model. The simplified loss computation formula is as follows:
\begin{equation}
\mathcal{L}_{diff} = \mathbb{E}_{t,x_0,\epsilon} \left[ ||\epsilon - \epsilon_\theta\left( \sqrt{\bar{\alpha}_t}x_0 + \sqrt{1 - \bar{\alpha}_t}\epsilon, t \right)||^2 \right],
\label{eq:4}
\end{equation}
where $\theta$ denotes the model parameters, $\epsilon$ denotes random noise, $x_0$ denotes the original data, $\bar{\alpha}_t$ is a parameter in the model, and $t$ is the time step. The expectation expression encapsulates the expectation value over all possible $t$, $x_0$, and $\epsilon$. Thus, the total loss of SPA-SVC is as follows:
\begin{equation}
\mathcal{L}_{total}=\mathcal{L}_{cyc}+\mathcal{L}_{diff}
\label{eq:5}
\end{equation}

As shown in Figure 2(b), the inference strategy of the model follows the approach of DDSP-SVC-Diff \cite{ddsp-svc2023}. We treat the DDSP model and the diffusion model collectively as the acoustic model. After extracting the $SSL_o$, $Vol_o$, and $F0_o$ features from the original audio and specifying the speaker ID, pitch-shifted $F0_s$ can be calculated by specifying the number of semitones for the pitch shifting. These features are then fed sequentially into the DDSP model and the diffusion model after passing through an embedding layer and being added together. The resulting $mel_s$ is then combined with $F0_s$ and fed into the pre-trained vocoder NSF-HiFiGAN \cite{openvpi2022nsf-hifigan} to obtain the desired timbre of the audio, thereby completing the SVC tasks.
\subsection{Modules}

    \subsubsection{Feature Extraction Model}
    \vspace{-0.1cm}
The Contentvec \cite{qian2022contentvec} is employed for extracting SSL features. This is an improved version of HuBERT \cite{hsu2021hubert} designed to better eliminate timbre information from speech encodings. Additionally, we utilize the pretrained RMVPE \cite{wei2023rmvpe} model in this study to extract the pitch, for its effectiveness and robustness.

    \subsubsection{DDSP Model}
    \vspace{-0.1cm}
To enable manipulation of energy, pitch, and timbre in vocal synthesis, we utilize the DDSP \cite{engel2020ddsp} model as our primary framework. DDSP integrates neural network methods with suitable feature representations, facilitating efficient learning and control of different sound aspects via controllable parameters. Its separation-based approach breaks down audio signals into key components like amplitude, frequency, and envelope, enabling accurate manipulation of energy, pitch, and timbre.

    \subsubsection{Diffusion Model}

Moreover, to improve the quality of vocal synthesis, the diffusion \cite{ho2020ddpm} model is utilized to process the output of DDSP. Integrating the diffusion model with DDSP enriches the description of sound characteristics by filling in any gaps, thereby boosting the performance of SVC. The strong performance of diffusion model in SVC tasks has been demonstrated in DiffSVC \cite{liu2021diffsvc} and CoMoSVC \cite{lu2024comosvc}.

    \subsubsection{NSF-HiFiGAN}
    \vspace{-0.1cm}
Furthermore, to achieve high-quality singing voices, NSF-HiFiGAN \cite{openvpi2022nsf-hifigan} is used as the vocoder in this study. Compared to HiFi-GAN \cite{kong2020hifigan}, incorporating the NSF \cite{wang2019nsf} module for modeling and processing audio signals allows for better capture and reconstruction of dynamic features and subtle structures within the audio. This enhances the quality and realism of singing voice synthesis.
\section{Experiments}
\label{sec:experiments}
\subsection{Datasets and Data Preprocessing}
To validate the effectiveness of the proposed method, we employed the M4Singer \cite{zhang2022m4singer} dataset for experimental training and inference. The M4Singer dataset is a large-scale high-quality Mandarin vocal corpus, comprising recordings from 20 professional singers covering 700 Mandarin pop songs and all four SATB types (i.e., soprano, alto, tenor, bass).
We randomly partitioned the data of each singer in the dataset at a ratio of approximately 9:1 to obtain training and testing sets. Specifically, we selected training sets of one male (Bass-3) and one female (Alto-1) singer. The training set of Alto-1 contains a total of 554 segments, while the testing set contains 61 segments. The training set of Bass-3 contains 967 segments, with 107 segments in the testing set. For actual inference, in addition to using the testing sets of these two singers, we further utilized the testing sets of soprano (Soprano-3) and tenor (Tenor-1) singers as unseen speakers' data, ensuring coverage across multiple vocal ranges. This allows us to validate that our model can generate high-quality target-domain voice even when provided with source-domain voice from different vocal ranges.

Our overall experimental hyperparameter settings largely follow the practices of DDSP-SVC-Diff \cite{ddsp-svc2023}. During the data preprocessing and feature extraction phase, audios were uniformly resampled to a sampling rate of 44.1 kHz, and segments shorter than 2 seconds were removed. We set the hop size to 512, extracting energy and 128-dimensional mel-spectrograms for loss computation. Additionally, we utilized a pretrained RMVPE \cite{wei2023rmvpe} to extract the pitch. We also employed a pretrained Contentvec \cite{qian2022contentvec} to extract 768-dimensional SSL features, with a sampling rate set at 16 kHz and hop size at 320.

\subsection{Experiment Setups}
    \label{sec:model_config}
In the model training process, we implemented the following settings: an initial learning rate of 0.00015, a batch size of 64, and a training step of 40k. We utilized the AdamW \cite{loshchilov2017adamw} optimizer, with $\beta_1 = 0.9$ and $\beta_2 = 0.999$. Furthermore, we used the StepLR scheduler, with its decay factor $\gamma = 0.5$. Similar to DDSP-SVC-Diff \cite{ddsp-svc2023}, we used DDSP \cite{engel2020ddsp} CombSubFast model and the WaveNet \cite{oord2016wavenet} decoder from Fastspeech2 \cite{ren2020fastspeech} in the diffusion model. WaveNet's input features were set to be dimension of 128, with 20 layers in the residual blocks, 512 output channels in the convolutional layers, and an encoder hidden layer size of 256. When using SSIM loss, we set default parameters $k_1=0.01$ and $k_2=0.03$. Moreover, we sequentially numbered the 20 singers of the M4Singer dataset from 1 to 20, according to alphabetical order, to serve as speaker ID inputs during training. Besides, for fair comparison, both our model and the DDSP-SVC-Diff model employed minor fundamental frequency random perturbations within the semitones of (-5, 5) range as the basis for data augmentation.

\begin{table*}[t]
  \captionsetup{justification=centering}
  \caption{MOS ratings for the public corpus M4Singer with the confidence interval 95\%}
  \label{tab:mos}
  \centering
  \rmfamily
  \begin{tabular*}{\textwidth}{@{\extracolsep{\fill}}lccccc}
    \toprule
    \textbf{Method} & \textbf{MOS-A} & \textbf{MOS-B} & \textbf{MOS-O} & \textbf{MOS-S} & \textbf{MOS}\\
    \midrule
    GT              & -             & -             & -             & -      & 4.30 ± 0.12       \\
    \midrule
    DDSP-SVC-Diff   & 3.95 ± 0.09 & 3.27 ± 0.07 & 3.70 ± 0.10 & 3.52 ± 0.08 & 3.61 ± 0.06      \\
    SPA-SVC-M       & 3.82 ± 0.09 & \textbf{3.60 ± 0.07} & 3.77 ± 0.09 & 3.65 ± 0.07 & 3.71 ± 0.06      \\
 \textbf{SPA-SVC (Ours)}      & \textbf{3.96 ± 0.09} & 3.56 ± 0.07 & \textbf{3.82 ± 0.09} & \textbf{3.69 ± 0.08} & \textbf{3.76 ± 0.06}       \\
    \bottomrule
  \end{tabular*}
\end{table*}

During model inference, we assigned speaker IDs as Alto-1 and Bass-3 and used audios from the testing sets as input for inference. To assess the model's performance, we conducted SVC with both identical and different source and target singers. We adjusted the pitch shifting based on the differences in pitch ranges between the source and target singers to maintain a balance between naturalness and pitch range. For instance, when the target singer was Alto-1 and the source singers were Alto-1, Bass-3, Soprano-3, and Tenor-1, we applied pitch shifting keys of 12, 12, 4, and 12, respectively. Similarly, when the target singer was Bass-3 and the source singers were Alto-1, Bass-3, Soprano-3, and Tenor-1, we applied pitch shifting keys of 4, 12, 4, and 4, respectively. Additionally, during model inference, we employed shallow diffusion with 100 steps on the diffusion model, which has been shown to enhance speech quality and accelerate inference speed \cite{liu2022diffsinger}.

\subsection{Experimental Results}
To assess the effectiveness of our method, we conducted Mean Opinion Score (MOS) evaluations on the proposed SPA-SVC, DDSP-SVC-Diff, and the SPA-SVC model with SSIM loss replaced by MSE loss. We ensured consistency in data preprocessing and model configuration. From the testing set, we randomly selected three audio segments each from four singers: Alto-1, Bass-3, Soprano-3, and Tenor-1, totaling 12 segments for model inference. Additionally, we collected 48 audio segments, both in original and shifted pitches, following the inference strategy outlined in Section \ref{sec:model_config}, with Alto-1 and Bass-3 as target timbres.

During the MOS rating process, 13 volunteers were invited to participate. They rated the quality of both real and generated audio on a 5-point scale after listening to audio samples from the target singers. Higher scores were given for naturalness and similarity to the target singer's timbre. Throughout the MOS scoring process, volunteers were unaware of which model generated the audio.

Table \ref{tab:mos} presents the MOS results along with the confidence interval 95\% on the M4Singer testing set. Here are explanations for the terms in the table: 
\begin{itemize}
    \item \textbf{MOS-A}: \textbf{MOS} for audio samples converted to \textbf{Alto-1} timbre, using both original and shifted pitch voices.
    \item \textbf{MOS-B}: \textbf{MOS} for audio samples converted to \textbf{Bass-3} timbre, using both original and shifted pitch voices.
    \item \textbf{MOS-O}: \textbf{MOS} for audio samples converted to either Alto-1 or Bass-3 timbre, using only \textbf{original pitch} voices.
    \item \textbf{MOS-S}: \textbf{MOS} for audio samples converted to either Alto-1 or Bass-3 timbre, using only \textbf{shifted pitch} voices.
    \item \textbf{GT}: Represents real audio input.
    \item \textbf{DDSP-SVC-Diff}: The baseline model.
    \item \textbf{SPA-SVC-M}: The SPA-SVC model with MSE loss.
    \item \textbf{SPA-SVC (Ours)}: The SPA-SVC model with SSIM loss.
\end{itemize}

\subsubsection{Comparison of MOS in general SVC scenarios}
\vspace{-0.2cm}
To evaluate the performance of the proposed SPA-SVC in general SVC scenarios, we compared SPA-SVC with DDSP-SVC-Diff at first. As shown in Table 1, both SPA-SVC-M and SPA-SVC achieve notably higher MOS scores than DDSP-SVC-Diff, with SPA-SVC outperforming DDSP-SVC-Diff by approximately 0.15, demonstrating superior performance of our training strategy in general SVC scenarios. Moreover, SPA-SVC exhibits a MOS improvement of around 0.05 compared to SPA-SVC-M, indicating that the introduction of SSIM loss in SVC domain enhances model's generation quality over conventional MSE loss. Additionally, in SVC scenarios without pitch shifting, the MOS-O of SPA-SVC still exceeds that of DDSP-SVC-Diff and SPA-SVC-M by approximately 0.12 and 0.05 respectively, further confirming the findings mentioned above.

\subsubsection{Comparison of MOS in cross-domain SVC scenarios}
\vspace{-0.2cm}
Furthermore, we assessed performance of SPA-SVC in cross-domain scenarios, with significant differences between target and source singer's vocal ranges or in cases of wide-ranging pitch shifting. We compared MOS-A, MOS-B, and MOS-S. Here, SPA-SVC's MOS-A slightly exceeds that of DDSP-SVC-Diff, indicating that the SPA-SVC model maintains its performance even in scenarios involving extreme high-pitched vocal generation. While MOS-B for SPA-SVC and SPA-SVC-M surpass DDSP-SVC-Diff by approximately 8.9\% and 10\% respectively, indicating that our training strategy significantly improves the performance of SVC in scenarios where bass singers perform songs in higher vocal ranges.  Comparing MOS-S, SPA-SVC consistently achieves superior performance, exceeding DDSP-SVC-Diff by approximately 0.17, once again highlighting its superiority in cross-domain scenarios.

\subsubsection{Comparison of Spectrograms}
\begin{figure}[t]
    \centering
        \includegraphics[width=\linewidth]{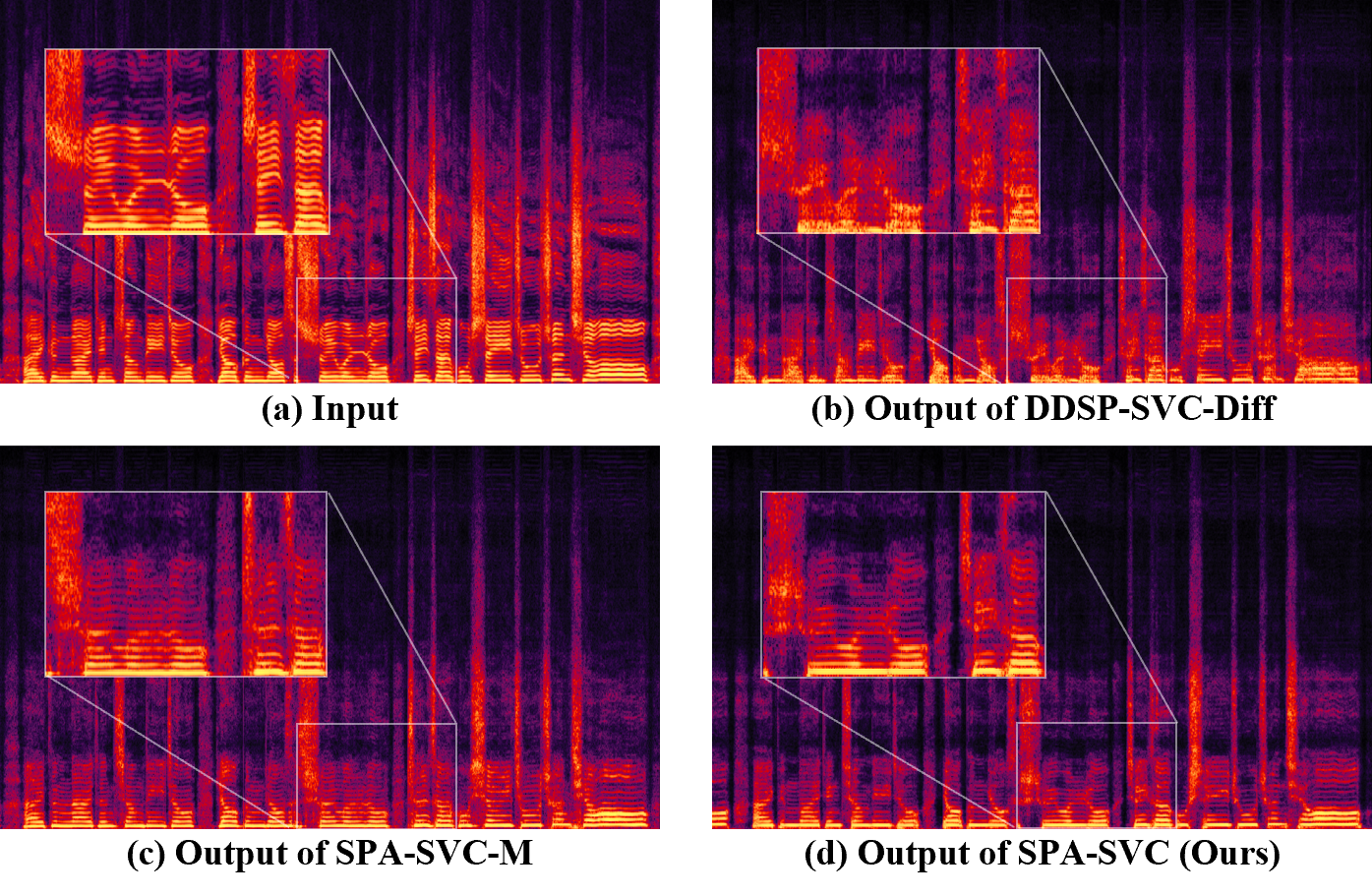}
        \caption{Comparison of spectrograms.}
        \label{fig:visualize}
\end{figure}

\vspace{-0.2cm}
To visually compare the actual performance of models in the cross-domain SVC scenario, we randomly selected a segment from Tenor-1 testing sets as the input and conducted SVC using Bass-3 as the target timbre. Subsequently, we obtained spectrograms of the waveforms generated by each model, as shown in Figure \ref{fig:visualize}. Comparing SPA-SVC with other models shows a clearer harmonic texture and less noise, which results in better sound quality.

\section{Conclusions}
\label{sec:conclusions}
In this study, we propose a novel self-supervised pitch augmentation method for singer voice conversion called SPA-SVC. By incorporating a cycle pitch shifting training strategy, our approach enables SVC models to overcome singer vocal range limitations, particularly in scenarios where the pitch of the source domain vocals significantly exceeds that of the target domain singer. Additionally, we introduced SSIM loss into SVC, allowing the SPA-SVC model to effectively learn vocal details and produce higher-quality vocals. Our experimental results demonstrate that, without requiring additional training data or increasing model parameters, our proposed SPA-SVC achieves optimal performance across most metrics. These results indicate that our method not only enhances the performance of in-domain SVC but also in cross-domain SVC tasks.

\section{Acknowledgement}
\label{sec:acknowledgement}
This study was supported by the National Natural Science Foundation of China (NSFC) (No. 62271083), the Key Project of the National Language Commission (No. ZDI145-81), the Fundamental Research Funds for the Central Universities (No. 2023RC73, 2023RC13), and was also supported in part by the Major Program of the National Social Science Fund of China (13\&ZD189).

\bibliographystyle{IEEEtran}
\bibliography{main}

\begin{thebibliography}{10}
\providecommand{\url}[1]{#1}
\csname url@samestyle\endcsname
\providecommand{\newblock}{\relax}
\providecommand{\bibinfo}[2]{#2}
\providecommand{\BIBentrySTDinterwordspacing}{\spaceskip=0pt\relax}
\providecommand{\BIBentryALTinterwordstretchfactor}{4}
\providecommand{\BIBentryALTinterwordspacing}{\spaceskip=\fontdimen2\font plus
\BIBentryALTinterwordstretchfactor\fontdimen3\font minus \fontdimen4\font\relax}
\providecommand{\BIBforeignlanguage}[2]{{%
\expandafter\ifx\csname l@#1\endcsname\relax
\typeout{** WARNING: IEEEtran.bst: No hyphenation pattern has been}%
\typeout{** loaded for the language `#1'. Using the pattern for}%
\typeout{** the default language instead.}%
\else
\language=\csname l@#1\endcsname
\fi
#2}}
\providecommand{\BIBdecl}{\relax}
\BIBdecl

\bibitem{baevski2020wav2vec2}
A.~Baevski, Y.~Zhou, A.~Mohamed, and M.~Auli, ``wav2vec 2.0: A framework for self-supervised learning of speech representations,'' \emph{Advances in neural information processing systems}, vol.~33, pp. 12\,449--12\,460, 2020.

\bibitem{hsu2021hubert}
W.-N. Hsu, B.~Bolte, Y.-H.~H. Tsai, K.~Lakhotia, R.~Salakhutdinov, and A.~Mohamed, ``Hubert: Self-supervised speech representation learning by masked prediction of hidden units,'' \emph{IEEE/ACM Transactions on Audio, Speech, and Language Processing}, vol.~29, pp. 3451--3460, 2021.

\bibitem{qian2022contentvec}
K.~Qian, Y.~Zhang, H.~Gao, J.~Ni, C.-I. Lai, D.~Cox, M.~Hasegawa-Johnson, and S.~Chang, ``Contentvec: An improved self-supervised speech representation by disentangling speakers,'' in \emph{International Conference on Machine Learning}.\hskip 1em plus 0.5em minus 0.4em\relax PMLR, 2022, pp. 18\,003--18\,017.

\bibitem{chen2022wavlm}
S.~Chen, C.~Wang, Z.~Chen, Y.~Wu, S.~Liu, Z.~Chen, J.~Li, N.~Kanda, T.~Yoshioka, X.~Xiao \emph{et~al.}, ``Wavlm: Large-scale self-supervised pre-training for full stack speech processing,'' \emph{IEEE Journal of Selected Topics in Signal Processing}, vol.~16, no.~6, pp. 1505--1518, 2022.

\bibitem{ddsp-svc2023}
Yxlllc, ``Real-time end-to-end singing voice conversion system based on ddsp,'' \url{https://github.com/yxlllc/DDSP-SVC}, 2023.

\bibitem{so-vits-svc2023}
SVC-Develop-Team, ``Softvc vits singing voice conversion,'' \url{https://github.com/svc-develop-team/so-vits-svc}, 2023.

\bibitem{liu2021diffsvc}
S.~Liu, Y.~Cao, D.~Su, and H.~Meng, ``Diffsvc: A diffusion probabilistic model for singing voice conversion,'' in \emph{2021 IEEE Automatic Speech Recognition and Understanding Workshop (ASRU)}.\hskip 1em plus 0.5em minus 0.4em\relax IEEE, 2021, pp. 741--748.

\bibitem{lu2024comosvc}
Y.~Lu, Z.~Ye, W.~Xue, X.~Tan, Q.~Liu, and Y.~Guo, ``Comosvc: Consistency model-based singing voice conversion,'' \emph{arXiv preprint arXiv:2401.01792}, 2024.

\bibitem{ho2020ddpm}
J.~Ho, A.~Jain, and P.~Abbeel, ``Denoising diffusion probabilistic models,'' \emph{Advances in neural information processing systems}, vol.~33, pp. 6840--6851, 2020.

\bibitem{kim2021conditional_vits}
J.~Kim, J.~Kong, and J.~Son, ``Conditional variational autoencoder with adversarial learning for end-to-end text-to-speech,'' in \emph{International Conference on Machine Learning}.\hskip 1em plus 0.5em minus 0.4em\relax PMLR, 2021, pp. 5530--5540.

\bibitem{xu2023mbtfnet}
W.~Xu, Z.~Chen, Z.~Tan, S.~Lv, R.~Han, W.~Zhou, W.~Zhao, and L.~Xie, ``Mbtfnet: Multi-band temporal-frequency neural network for singing voice enhancement,'' in \emph{2023 IEEE Automatic Speech Recognition and Understanding Workshop (ASRU)}.\hskip 1em plus 0.5em minus 0.4em\relax IEEE, 2023, pp. 1--8.

\bibitem{liu2022voicefixer}
H.~Liu, X.~Liu, Q.~Kong, Q.~Tian, Y.~Zhao, D.~Wang, C.~Huang, and Y.~Wang, ``Voicefixer: A unified framework for high-fidelity speech restoration,'' \emph{arXiv preprint arXiv:2204.05841}, 2022.

\bibitem{guo2022singaug}
S.~Guo, J.~Shi, T.~Qian, S.~Watanabe, and Q.~Jin, ``Singaug: Data augmentation for singing voice synthesis with cycle-consistent training strategy,'' \emph{arXiv preprint arXiv:2203.17001}, 2022.

\bibitem{liu2021fastsvc_cross}
S.~Liu, Y.~Cao, N.~Hu, D.~Su, and H.~Meng, ``Fastsvc: Fast cross-domain singing voice conversion with feature-wise linear modulation,'' in \emph{2021 ieee international conference on multimedia and expo (icme)}.\hskip 1em plus 0.5em minus 0.4em\relax IEEE, 2021, pp. 1--6.

\bibitem{zhou2022hifisvc_cross}
Y.~Zhou and X.~Lu, ``Hifi-svc: Fast high fidelity cross-domain singing voice conversion,'' in \emph{ICASSP 2022-2022 IEEE International Conference on Acoustics, Speech and Signal Processing (ICASSP)}.\hskip 1em plus 0.5em minus 0.4em\relax IEEE, 2022, pp. 6667--6671.

\bibitem{wang2004ssim}
Z.~Wang, A.~C. Bovik, H.~R. Sheikh, and E.~P. Simoncelli, ``Image quality assessment: from error visibility to structural similarity,'' \emph{IEEE transactions on image processing}, vol.~13, no.~4, pp. 600--612, 2004.

\bibitem{zhou2023vitssvcc}
Y.~Zhou, M.~Chen, Y.~Lei, J.~Zhu, and W.~Zhao, ``Vits-based singing voice conversion system with dspgan post-processing for svcc2023,'' in \emph{2023 IEEE Automatic Speech Recognition and Understanding Workshop (ASRU)}.\hskip 1em plus 0.5em minus 0.4em\relax IEEE, 2023, pp. 1--8.

\bibitem{peng2023singing_vits_opera}
Z.~Peng, J.~Wu, and Y.~Li, ``Singing voice conversion between popular music and chinese opera based on vits,'' in \emph{2023 IEEE Intl Conf on Dependable, Autonomic and Secure Computing, Intl Conf on Pervasive Intelligence and Computing, Intl Conf on Cloud and Big Data Computing, Intl Conf on Cyber Science and Technology Congress (DASC/PiCom/CBDCom/CyberSciTech)}.\hskip 1em plus 0.5em minus 0.4em\relax IEEE, 2023, pp. 0999--1003.

\bibitem{openvpi2022nsf-hifigan}
OpenVPI, ``Diffsinger (openvpi maintained version),'' \url{https://github.com/openvpi/DiffSinger}, 2022.

\bibitem{wei2023rmvpe}
H.~Wei, X.~Cao, T.~Dan, and Y.~Chen, ``Rmvpe: A robust model for vocal pitch estimation in polyphonic music,'' \emph{arXiv preprint arXiv:2306.15412}, 2023.

\bibitem{engel2020ddsp}
J.~Engel, L.~Hantrakul, C.~Gu, and A.~Roberts, ``Ddsp: Differentiable digital signal processing,'' \emph{arXiv preprint arXiv:2001.04643}, 2020.

\bibitem{kong2020hifigan}
J.~Kong, J.~Kim, and J.~Bae, ``Hifi-gan: Generative adversarial networks for efficient and high fidelity speech synthesis,'' \emph{Advances in neural information processing systems}, vol.~33, pp. 17\,022--17\,033, 2020.

\bibitem{wang2019nsf}
X.~Wang, S.~Takaki, and J.~Yamagishi, ``Neural source-filter-based waveform model for statistical parametric speech synthesis,'' in \emph{ICASSP 2019-2019 IEEE International Conference on Acoustics, Speech and Signal Processing (ICASSP)}.\hskip 1em plus 0.5em minus 0.4em\relax IEEE, 2019, pp. 5916--5920.

\bibitem{zhang2022m4singer}
L.~Zhang, R.~Li, S.~Wang \emph{et~al.}, ``M4singer: A multi-style, multi-singer and musical score provided mandarin singing corpus,'' \emph{Advances in Neural Information Processing Systems}, vol.~35, pp. 6914--6926, 2022.

\bibitem{loshchilov2017adamw}
I.~Loshchilov and F.~Hutter, ``Decoupled weight decay regularization,'' \emph{arXiv preprint arXiv:1711.05101}, 2017.

\bibitem{oord2016wavenet}
A.~v.~d. Oord, S.~Dieleman, H.~Zen, K.~Simonyan, O.~Vinyals, A.~Graves, N.~Kalchbrenner, A.~Senior, and K.~Kavukcuoglu, ``Wavenet: A generative model for raw audio,'' \emph{arXiv preprint arXiv:1609.03499}, 2016.

\bibitem{ren2020fastspeech}
Y.~Ren, C.~Hu, X.~Tan, T.~Qin, S.~Zhao, Z.~Zhao, and T.-Y. Liu, ``Fastspeech 2: Fast and high-quality end-to-end text to speech,'' \emph{arXiv preprint arXiv:2006.04558}, 2020.

\bibitem{liu2022diffsinger}
J.~Liu, C.~Li, Y.~Ren, F.~Chen, and Z.~Zhao, ``Diffsinger: Singing voice synthesis via shallow diffusion mechanism,'' in \emph{Proceedings of the AAAI conference on artificial intelligence}, vol.~36, no.~10, 2022, pp. 11\,020--11\,028.

\end{thebibliography}

\end{document}